\documentstyle[aps]{revtex}
\tightenlines

\begin{document}
\draft

\title{Ferromagnetic Fixed Point of the Kondo Model in a Luttinger Liquid}
\author{Mun Dae Kim$^{a,b}$, Chul Koo Kim$^{a,b}$, Kyun Nahm$^c$
        and Chang-Mo Ryu$^d$}
\address{
        $^a$Institute of Physics and Applied Physics, Yonsei University,
          Seoul 120-749, Korea\\
        $^b$Center for Strongly Correlated Materials Research,
          Seoul National University, Seoul 151-742,Korea\\
        $^c$Department of Physics, Yonsei University, Wonju 220-710,Korea\\
        $^d$Department of Physics, Pohang University of Science and
          Technology, Pohang 790-784, Korea
        }
\maketitle

\begin{abstract}
The Kondo effect in a Luttinger liquid is studied using the
renormalization group method. By renormalizing the boson fields,
scaling equations to the second order for an arbitrary Luttinger
interaction are  obtained. For the ferromagnetic Kondo coupling, a spin bound
state(triplet) can be realized without invoking a nearest neighbor
spin interaction in agreement with the recent Bethe ansatz
calculation. The scaling theory in the presence of the scalar
potential shows that there is no interplay between the magnetic
and non-magnetic interaction. Also a study on the crossover
behavior of the Kondo temperature between the exponential and the
power law type is presented.
\end{abstract}


\section{Introduction}

The Kondo problem in one-dimensional quantum system has attracted
substantial interest in connection with the recent rapid
development of the nanofabrication technology. In one-dimension,
interacting electron system is described by the Tomonaga-Luttinger
liquid theory\cite{Hal,Emery,Voit,Gogolin,Sch}, whose low- energy
excitations are not quasiparticles but collective charge and spin
density fluctuations. The magnetic impurity effect in this
non-Fermi liquid was first studied  by Lee and Toner using a
scaling analysis on the kink gas action \cite{LT}. They obtained
the scaling equation to the first order of coupling constant in
the weak coupling regime. Furusaki and Nagaosa(FN) extended the
study and obtained a set of scaling equation  up to the second
order using the poor man's scaling theory\cite{FN}. In their
study, FN proposed an interesting conjecture that even
ferromagnetic Kondo impurity as well as antiferromagnetic one
would be completely screened. In the antiferromagnetic coupling
case, the coupling constant flows to a strong coupling regime and,
thus, the magnetic impurity and a conduction electron form a
singlet. But, in the ferromagnetic case, situation is not
clear whether it flows into a strong coupling regime or not. In
order to clarify the situation, FN considered coupling of impurity
spins not only with the same site electrons but also with the
nearest neighbor conduction electrons. In this picture, the
impurity and three conduction electrons form a singlet
composite. However, a recent Bethe ansatz(BA) result by the Wang
and Voit showed that spins form a triplet for the ferromagnetic
coupling case\cite{WV}. These conflicting results call for a more detailed
renormalization group(RG) analysis to clarify the physical
situation.

In this paper, we carry out an RG analysis which goes beyond the poor man's scaling scheme in the
presence of the Kondo interaction. We bosonize the Kondo interaction term using the Abelian
bosonization. Carrying out a full RG calculation for the second order cumulant, we obtain a set
of general scaling equations valid for an arbitrary strength of the Luttinger interaction. We
show that the present RG calculation confirms the calculation drawn from the BA calculation using
an open boundary condition.

Another recent BA study of the same problem by Wang and Eckern
showed that there is a competition between the Kondo coupling and
the impurity potential\cite{WE}. When the impurity potential is
dominant, the system is shown to flow to a weak coupling fixed
point. However, when the magnetic interaction is dominant, a spin
complex is shown to be formed. In order to investigate this
problem, we bosonize the potential scattering term and perform a
scaling calculation. The result shows that the magnetic
interaction and the potential scattering do not interplay in the
scaling procedure and flow independently.

In the first order scaling, the Kondo temperature in the Luttinger
can be calculated analytically to show a power-law behavior in
contrast to the exponential one of the conventional Kondo model
\cite{LT}. However, it is not possible to show the crossover from
the Fermi to the Luttinger liquid in the first order. In this
paper, we study the crossover behavior of the Kondo temperature
from the exponential to the power-law type as a function of  the
Luttinger interaction, by solving the RG equations.

\section{Bosonization of the Kondo Hamiltonian and the Partition Function}

Introducing the  two phase fields,
\begin{eqnarray}
\phi_\nu &=& -\frac{i\pi}{L}\sum_{p\neq 0}\frac{e^{-\alpha |p|/2-ipx}}{p}
\left(\nu_+(p)+\nu_-(p)\right)-\left(N_{+\nu}+N_{-\nu}\right)\frac{\pi x}{L}, \nonumber \\
\theta_\nu &=& \frac{i \pi}{L}\sum_{p \neq0}\frac{e^{-\alpha |p|/2-ipx}}{p}
\left(\nu_+(p)-\nu_-(p)\right)+\left(N_{+\nu}-N_{-\nu}\right)\frac{\pi x}{L},
\end{eqnarray}
the one-dimensional model Hamiltonian with the forward electron-electron scattering is written by
\begin{eqnarray}
H=\frac{1}{2\pi}\sum_\nu\int dx\left(v_\nu \eta_\nu\pi^2\pi_\nu^2(x)
 +\frac{v_\nu}{\eta_\nu}
  \left(\frac{\partial\phi_\nu(x)}{\partial x}\right)^2\right),
\end{eqnarray}
where $\nu=\rho$ or $\sigma$, $+(-)$ means right(left) going mode, and
the parameter $v_\nu$ and $\eta_\nu$ are given by
\begin{eqnarray}
v_\nu&=&\sqrt{\left(v_F+\frac{g_{4\nu}}{\pi}\right)^2
              -\left(\frac{g_{2\nu}}{\pi}\right)^2},
\hspace{5mm}
\eta_\nu=\sqrt{\frac{v_F+\frac{g_{4\nu}-g_{2\nu}}{\pi}}
                    {v_F+\frac{g_{4\nu}+g_{2\nu}}{\pi}}
              }.
\end{eqnarray}
Here, $g_2$ and $g_4$ are the dimensionless coupling parameters\cite{Voit}. The fields $\phi_\nu$ and
$\pi_\nu$ satisfy the canonical boson commutation relation;

\begin{eqnarray}
\left[\phi_{\nu}(x),\pi_{\nu'}(x')\right]
     &=&i\delta_{\nu\nu'}\delta(x-x'),\\
\left[\phi_{\nu}(x),\theta_{\nu'}(x')\right]
     &=&i\frac{\pi}{2}\delta_{\nu\nu'}\mbox{sign}(x'-x),
\hspace{5mm}
\theta_\nu(x)=\pi\int_{-\infty}^x \pi_\nu(y) dy.\nonumber
\end{eqnarray}

Alternatively, this Hamiltonian can also be obtained directly from the fermion
representation of the one dimensional interacting electron system using
\begin{eqnarray}
\psi_{rs}(x)=\lim_{\alpha \rightarrow 0} \frac{e^{ir \left(k_F-\frac{\pi}{L}\right)x}}
     {\sqrt{2\pi\alpha}} \eta_{rs} e^{-\frac{i}{\sqrt{2}}\left(r\phi_\rho(x)-\theta_\rho(x)
     +s\left(r\phi_\sigma(x)-\theta_\sigma(x)\right)\right)},
\end{eqnarray}
where $\eta_{rs}$ is the Majorana fermion operator which satisfies the
following relations\cite{Sch};

\begin{eqnarray}
[\eta_{rs},\eta_{r's'}]&=&2\delta_{rr'}\delta_{ss'} \\
\eta_{+\downarrow}\eta_{+\uparrow}
=\eta_{-\downarrow}\eta_{-\uparrow},\hspace{5mm}
\eta_{+\downarrow}\eta_{-\uparrow}
&=&-\eta_{-\downarrow}\eta_{+\uparrow},\hspace{5mm}
\eta_{+\uparrow}\eta_{-\uparrow}
=-\eta_{+\downarrow}\eta_{-\downarrow} \hspace{5mm}
\cdots\nonumber
\end{eqnarray}

   If a single magnetic impurity is introduced in one-dimensional interacting electron
system, the Kondo interaction term is given by
\begin{eqnarray}
H_K&=&J\vec{S}\cdot\vec{s}(0) \nonumber \\
   &=&J_zS_zs_z(0)+\frac{1}{2}J_\perp(S_+s_-(0) + S_-s_+(0)),
\end{eqnarray}
where $\vec{s}=\frac{1}{2}\sum_{rr' \sigma \sigma'}\psi_{r
\sigma}^\dagger \vec{\sigma}_{\sigma \sigma'}\psi_{r' \sigma'}$
and $s_\pm=s_x \pm is_y$. Using the relations of the Majorana
fermions, Eq.(6), and the bosonization formula of the fermion
operators, we obtain the bosonized Kondo Hamiltonian\cite{EK},
\begin{eqnarray}
H_K&=&J\vec{S} \cdot \vec{s} \nonumber \\
   &=&\frac{S_z}{2\pi\alpha}(J_{zF}\alpha\partial_x\sqrt{2}\phi_\sigma(0)
     +2iJ_{zB}\eta_{+\uparrow}\eta_{-\uparrow}\sin(\sqrt{2}\phi_\rho(0))\cos(\sqrt{2}\phi_\sigma(0)))
      \nonumber \\
   &+&\frac{S_+}{2\pi\alpha}e^{\sqrt{2} i \theta_\sigma(0)}
     (J_{\perp F}\eta_{+\downarrow}\eta_{+\uparrow}\cos(\sqrt{2}\phi_\sigma(0))
    +iJ_{\perp B}\eta_{+\downarrow}\eta_{-\uparrow}\sin(\sqrt{2}\phi_\rho(0)))\\
   &+&\frac{S_-}{2\pi\alpha}e^{-\sqrt{2} i \theta_\sigma(0)}
     (J_{\perp F}\eta_{+\uparrow}\eta_{+\downarrow}\cos(\sqrt{2}\phi_\sigma(0))
    -iJ_{\perp B}\eta_{-\uparrow}\eta_{+\downarrow}\sin(\sqrt{2}\phi_\rho(0))). \nonumber
\end{eqnarray}
The partition function of the system at temperature $T=1/\beta$ is
\begin{eqnarray}
Z=\int& &D\phi_\rho D\phi_\sigma D\theta_\rho D\theta_\sigma e^{-S},\\
S& &=\int dx \int d\tau(L_0+L_K), \nonumber \\
L_0& &=\sum_{\nu=\rho,\sigma}\left(i\partial_\tau\phi_\nu(x,\tau) \pi_\nu(x,\tau)
    +\frac{1}{2\pi}\left(v_\nu \eta_\nu\pi^2\pi_\nu^2(x,\tau)+\frac{v_\nu}{\eta_\nu}
     \left(\frac{\partial\phi_\nu(x,\tau)}{\partial x}\right)^2\right)\right), \nonumber \\
L_K& &=H_K(\phi_\nu(0,\tau),\theta_\nu(0,\tau)), \nonumber
\end{eqnarray}
where the integration is over the bosonic fields $\phi_\nu(x,\tau)$
and $\theta_\nu(x,\tau)$ with imaginary time $\tau$ running from 0
to $\beta$.

\section{Renormalization Analysis}

First, we divide the phase fields $\phi_\nu$ into slow and fast
mode:
\begin{eqnarray}
\phi_\nu(\tau)=\phi_{\nu s}(\tau)&+&\phi_{\nu f}(\tau) \nonumber \\
\phi_{\nu s}(\tau)&=&\frac{1}{\beta}\sum_{|\omega_n|<\mu}
                     \tilde{\phi_\nu}(\omega)e^{-i\omega\tau}\\
\phi_{\nu f}(\tau)&=&\frac{1}{\beta}\sum_{\mu<|\omega_n|<\lambda}
                     \tilde{\phi_\nu}(\omega)e^{-i\omega\tau}.\nonumber
\end{eqnarray}
The average over the fast mode of the partition function is carried out, using the cumulant expansion,
\begin{eqnarray}
Z &=& Z_0<e^{-S_K}>_0 \nonumber \\
  &=& Z_0\int D\phi_{\nu s}D\theta_{\nu s}
  e^{-S_0(\phi_{\nu s},\theta_{\nu s})}
  e^{-<S_K>_0^f+\frac{1}{2}(<{S_K}^2>_0^f-{<S_K>_0^f}^2)+\cdots},
\end{eqnarray}
where $Z_0=\int D\phi_\nu D\theta_\nu e^{-S_0}$, f indicates an average over the fast mode which
will be omitted hereafter, and $<\cdot\cdot\cdot>_0$ represents an average over $Z_0$.

First, we consider the first order forward longitudinal scattering
term,
\begin{eqnarray}
<\int d\tau J_{zF}\alpha\partial_x\sqrt{2}\phi_\sigma(\tau)>
=\int d\tau J_{zF}\alpha(\partial_x\sqrt{2}\phi_{\sigma s}(\tau)
+<\partial_x\sqrt{2}\phi_{\sigma f}(\tau)>).
\end{eqnarray}
The second term in the right side vanishes because it is an
average of an odd function. We, thus, conclude that $\delta
J_{zF}=0$ in the first order. The backward longitudinal scattering
part is scaled as follows,
\begin{eqnarray}
\int d\tau J_{zB}\eta_{+\uparrow}\eta_{-\uparrow}
<\sin(\sqrt{2}\phi_\rho(\tau)><\cos(\sqrt{2}\phi_\sigma(\tau))> \nonumber \\
=\left(\frac{\mu}{\lambda}\right)^{\frac{\eta_\rho}{2}+\frac{\eta_\sigma}{2}}\int d\tau
  J_{zB}\eta_{+\uparrow}\eta_{-\uparrow}
  \sin(\sqrt{2}\phi_{\rho s}(\tau)\cos(\sqrt{2}\phi_{\sigma s}(\tau)).
\end{eqnarray}
In the above, we utilized the fact that the charge and spin
degrees are separated. The rescaling procedure,
\begin{eqnarray}
J_{zB}(\mu)=\left(\frac{\mu}{\lambda}\right)^{\frac{\eta_\rho+\eta_\sigma}{2}-1}J_{zB}(\lambda)
\end{eqnarray}
gives
\begin{eqnarray}
\frac{\delta J_{zB}}{J_{zB}}=\left(\frac{\eta_\rho+\eta_\sigma}{2}-1\right)\frac{\delta\lambda}{\lambda},
\end{eqnarray}
where $\mu=\lambda+\delta\lambda$ and
$\delta l=-\frac{\delta\lambda}{\lambda}=-\delta\ln{\lambda}.$
Thus, we have in  the first order
\begin{eqnarray}
\frac{dJ_{zF}}{dl}&=&0,\nonumber\\
\frac{dJ_{zB}}{dl}&=&\left(1-\frac{\eta_\rho+\eta_\sigma}{2}\right)J_{zB}.
\end{eqnarray}
The scaling equations for the other scattering terms can be similarly
obtained;
\begin{eqnarray}
\frac{dJ_{\perp F}}{dl}
&=&\left[1-\left(\frac{1}{2\eta_\sigma}+\frac{\eta_\sigma}{2}\right)\right]
   J_{\perp F}, \nonumber \\
\frac{dJ_{\perp B}}{dl}
&=&\left[1-\left(\frac{1}{2\eta_\sigma}+\frac{\eta_\rho}{2}\right)\right]
   J_{\perp B}.
\end{eqnarray}
These equations are in agreement with those of Lee and Toner \cite{LT}.

The second order cumulant is given by $-\frac{1}{2}(<S_K^2>-<S_K>^2)$, where $<S_K>^2$ term is to
eliminate unconnected diagrams. We consider one of the $J_{zB}J_{\perp F}$ terms which is given
by
\begin{eqnarray}
   \int d\tau\int d\tau'& &\frac{S_z S_+}{(2\pi\alpha)^2}
   J_{zB}2i\eta_{+\uparrow}\eta_{-\uparrow}
   J_{\perp F}\eta_{+\downarrow}\eta_{+\uparrow} \nonumber \\
& &\left(<\sin(\sqrt{2}\phi_\rho(\tau))\cos(\sqrt{2}\phi_\sigma(\tau))
   e^{\sqrt{2}i\theta_\sigma(\tau')}cos(\sqrt{2}\phi_\sigma(\tau'))>\right.\\
& &\left.-<\sin(\sqrt{2}\phi_\rho(\tau))\cos(\sqrt{2}\phi_\sigma(\tau))>
   <e^{\sqrt{2}i\theta_\sigma(\tau')}\cos(\sqrt{2}\phi_\sigma(\tau'))>\right). \nonumber
\end{eqnarray}
In order to evaluate this expression, we need the two point correlation function
\cite{FZ,GS,Stone,Nomura}
\begin{eqnarray}
G(x,\tau)&=&<\phi(x,\tau)\phi(0,0)>\\
         &=&\int\frac{dq}{2\pi}\int\frac{d\omega}{2\pi}e^{-iqx}e^{i\omega\tau}
            \frac{\pi}{\frac{1}{v\eta}\omega^2+\frac{v}{\eta}q^2} \nonumber \\
G(\tau)\equiv && G(0,\tau) \nonumber \\
         &=&\left\{
            \begin{array}{ll}
                  \frac{\eta}{2}K_0(\mu\tau)            & \mbox{for $\lambda\tau >> 1 $} \\
                   \frac{\eta}{2}\ln{\frac{\lambda}{\mu}} & \mbox{for $\lambda\tau << 1,$}
           \end{array}
            \right. \nonumber
\end{eqnarray}
where $K_0$ is the modified Bessel function of the second kind.
$G(\tau)$ decays exponentially with the renormalized lattice
spacing $\frac{1}{\mu}$, and decreases logarithmically for small
$\lambda\tau$. Thus, we regard $G(\tau)$ short ranged and, thus,
expand the cosine terms around zero. It is known that the higher
harmonics $\cos(2\sqrt{2}\phi(0))$ is irrelevant and
$((\partial\phi(\tau)/\partial\tau)|_{\tau=0})^2$ terms which is
the most relevant term in the expansion can also be shown
irrelevant by the power counting \cite{FZ}. Short time cut off,
$\tau_0 \sim \alpha/v_F$, merely introduces an overall constant
which does not affect the flow of the parameter. Therefore Eq.(18)
is reduced to
\begin{eqnarray}
\int d\tau\frac{\alpha}{v_F}\frac{1}{(2\pi\alpha)^2}
\frac{1}{2}S_+J_{zB} J_{\perp F}(-i\eta_{+\downarrow}\eta_{-\uparrow})\left
(\frac{\mu}{\lambda}\right)^{\frac{\eta_\rho}{2}+\frac{1}{2\eta_\sigma}+\eta_\sigma}
\left(\left(\frac{\mu}{\lambda}\right)^{-\eta_\sigma}-1\right)
e^{\sqrt{2}i\theta_\sigma(\tau)}\sin(\sqrt{2}\phi_\rho(\tau)).
\end{eqnarray}
  Collecting other terms of the second order cumulant and rescaling as before, we have
for the transverse backward part,
\begin{eqnarray}
\int d\tau\frac{S_+}{2\pi\alpha}i\eta_{+\downarrow}\eta_{-\uparrow}
\eta_\sigma J_{zB} J_{\perp F} \frac{d\lambda}{\lambda}
e^{\sqrt{2}i\theta_{\sigma s}(\tau)}\sin(\sqrt{2}\phi_{\rho s}(\tau)),
\end{eqnarray}
which renormalize the coupling constant $J_{\perp B}$ of the Kondo
term. Similarly, we have for the transverse forward scattering
part which renormalizes the coupling constant $J_{\perp F},$
\begin{eqnarray}
\int d\tau\frac{S_+}{2\pi\alpha}\eta_{+\downarrow}\eta_{+\uparrow}
\eta_\rho J_{zB} J_{\perp B} \frac{d\lambda}{\lambda}
e^{\sqrt{2}i\theta_\sigma(\tau)}\cos(\sqrt{2}\phi_\sigma(\tau)).
\end{eqnarray}
The same scaling process on the $S_-$ terms gives the same renormalization for both the $J_{\perp
B}$ and $J_{\perp F}$.

  However, the scaling process involving the descendant field terms,
$J_{zF}\alpha\partial_x\sqrt{2}\phi_\sigma(\tau)$, is somewhat
different. One example of such a term is the second order cumulant
for the transverse forward scattering, which is given by
\begin{eqnarray}
\frac{S_z S_+}{(2\pi\alpha)^2}J_{zF} J_{\perp
F}\eta_{+\downarrow}\eta_{+\uparrow}
\left(<\alpha\partial_x\sqrt{2}\phi_\sigma(\tau)e^{i\sqrt{2}\theta_\sigma(\tau')}
\cos(\sqrt{2}\phi_\sigma(\tau')>\right. \nonumber \\
 \left.-<\alpha\partial_x\sqrt{2}\phi_\sigma(\tau)>
 <e^{i\sqrt{2}\theta_\sigma(\tau')}\cos(\sqrt{2}\phi_\sigma(\tau'))>\right)\\
=\frac{S_z S_+}{(2\pi\alpha)^2}J_{zF}J_{\perp F}\eta_{+\downarrow}\eta_{+\uparrow}
<\alpha\partial_x\sqrt{2}\phi_{\sigma f}(\tau)e^{i\sqrt{2}\theta_\sigma(\tau')}
\cos(\sqrt{2}\phi_\sigma(\tau'))>. \nonumber
\end{eqnarray}
Here, we note that
\begin{eqnarray}
&&<\alpha\partial_x\sqrt{2}\phi_{\sigma f}(\tau) e^{i\sqrt{2}\theta_\sigma(\tau')}
e^{i\sqrt{2}\phi_\sigma(\tau')}> \nonumber \\
&=&\lim_{\epsilon\rightarrow 0}\frac{1}{i\epsilon}\frac{\partial}{\partial x}
<\alpha e^{i\epsilon\sqrt{2}\phi_{\sigma f}(x,\tau)}e^{i\sqrt{2}\theta_\sigma(0,\tau')}
e^{i\sqrt{2}\phi_\sigma(0,\tau')}>|_{x=0}.
\end{eqnarray}
Using $e^{A+B}=e^Ae^Be^{-\frac{1}{2}[A,B]},$ the above expression takes a form,
\begin{eqnarray}
\frac{\alpha}{i}(2\partial_x G_{\phi_{\sigma f}}(x,\tau)
-2\partial_x<\phi_{\sigma f}(x,\tau)\theta_{\sigma f}(0,0)>)|_{x=0}
e^{-G_{\theta_{\sigma f}}(0,0)-G_{\theta_{\phi f}}(0,0)}.
\end{eqnarray}
In order to calculate $\partial_x<\phi_{\sigma
f}(x,\tau)\theta_{\sigma f}(0,0)>$, we use the relations between
density field, $\phi_\sigma$, and current field, $\theta_\sigma$
\cite{Voit,Nomura};
\begin{eqnarray}
-\frac{i}{\eta_\sigma}\frac{\partial\phi_\sigma}{\partial(v_\sigma
\tau)} &=&\frac{\partial\theta_\sigma}{\partial x}, \nonumber \\
-i\eta_\sigma\frac{\partial\theta_\sigma}{\partial(v_\sigma \tau)}
&=&\frac{\partial\phi_\sigma}{\partial x}.
\end{eqnarray}
 Then, we have
\begin{eqnarray}
\alpha\partial_x<\phi_{\sigma f}(x,\tau)\theta_{\sigma f}(0,0)>
&=&\alpha\eta_\sigma\frac{\partial}{\partial(v_\sigma \tau)}
<\theta_{\sigma f}(x,\tau)\theta_{\sigma f}(0,0)> \nonumber \\
&=&\frac{1}{2}\frac{d\lambda}{\lambda},
\end{eqnarray}
using Eq.(19) with $\frac{1}{\eta}\rightarrow\eta$ and the short-ranged nature of the correlation
function $G(\tau)$. Substituting this result into Eq.(23), we obtain
\begin{eqnarray}
\frac{1}{2\pi\alpha} S_+\eta_{+\downarrow}\eta_{+\uparrow}
\left(\frac{1}{2\pi v_F} J_{zF} J_{\perp F} \frac{d\lambda}{\lambda}\right)
e^{i\sqrt{2}\theta_\sigma(\tau)}\cos(\sqrt{2}\phi_\sigma(\tau)) \nonumber \\
+\frac{1}{2\pi\alpha}S_+\eta_{+\downarrow}\eta_{-\uparrow}
\left(\frac{1}{2\pi v_F} J_{zF} J_{\perp B} \frac{d\lambda}{\lambda}\right)
e^{i\sqrt{2}\theta_\sigma(\tau)}\sin(\sqrt{2}\phi_\rho(\tau)),
\end{eqnarray}
for the renormalization of $J_{\perp F}$ and $J_{\perp B}$.

The longitudinal scattering parameter is scaled
by consecutive transverse scattering processes. The forward
scattering part which contains $J_{\perp F}^2$ is
\begin{eqnarray}
\frac{S_+S_-}{(2\pi\alpha)^2}J_{\perp F}^2
\eta_{+\downarrow}\eta_{+\uparrow}\eta_{+\uparrow}\eta_{+\downarrow}
\left[<e^{\sqrt{2}i\theta_\sigma(\tau)}\cos(\sqrt{2}\phi_\sigma(\tau))
e^{-\sqrt{2}i\theta_\sigma(\tau')}\cos(\sqrt{2}\phi_\sigma(\tau'))>\right.\nonumber \\
\left.-<e^{\sqrt{2}i\theta_\sigma(\tau)}\cos(\sqrt{2}\phi_\sigma(\tau))>
<e^{-\sqrt{2}i\theta_\sigma(\tau')}\cos(\sqrt{2}\phi_\sigma(\tau'))>\right].
\end{eqnarray}
The typical relevant term is given by
\begin{eqnarray}
e^{\sqrt{2}i\theta_\sigma(\tau)}e^{-\sqrt{2}i\theta_\sigma(\tau')}
e^{-\sqrt{2}i\phi_\sigma(\tau)}e^{\sqrt{2}i\phi_\sigma(\tau')}.
\end{eqnarray}
Other terms containing higher harmonics, $e^{2\sqrt{2}i\theta_\sigma(\tau)}$ or
$e^{2\sqrt{2}i\phi_\sigma(\tau)}$ are  irrelevant \cite{FZ,Wiegmann}. Separating  the fields into
the fast and slow mode, we average on the fast mode to obtain a short range correlation and expand
the slow mode to obtain
\begin{eqnarray}
1+i\sqrt{2}\partial_\tau\theta_{\sigma s}(\tau)|_{\tau=0}\tau
+i\sqrt{2}\partial_\tau\phi_{\sigma s}(\tau)|_{\tau=0}\tau+\cdots .
\end{eqnarray}
We note that $\partial_\tau\theta_{\sigma s}(\tau)$ gives the longitudinal scattering
contribution, $\partial_x\phi_{\sigma s}$, through Eq.(26), while $\partial_\tau\phi_{\sigma s}$
term is canceled by other terms. Including the contributions from the $S_-S_+$ term, we obtain
\begin{eqnarray}
\int d\tau\left[\frac{1}{2\pi\alpha}\frac{S_z}{2\pi v_F}\frac{1}{\eta_\sigma}
\left(\frac{1}{2\eta_\sigma}+\frac{\eta_\sigma}{2}\right)
J_{\perp F}^2\frac{d\lambda}{\lambda}\alpha\partial_x\sqrt{2}\phi_{\sigma s}(\tau)\right. \nonumber\\
\left.+\frac{1}{2\pi\alpha}\frac{S_z}{2\pi v_F}\frac{1}{\eta_\sigma}
\left(\frac{1}{2\eta_\sigma}+\frac{\eta_\rho}{2}\right)J_{\perp B}^2
\frac{d\lambda}{\lambda}\alpha\partial\sqrt{2}\phi_{\sigma s}(\tau)\right],
\end{eqnarray}
for the $J_{\perp B}^2$ contribution to the $J_{zF}$
renormalization. Consecutive transverse scattering also
renormalizes the longitudinal backward term, $J_{zB},$ similarly;
\begin{eqnarray}
\int d\tau\frac{S_z}{2\pi\alpha}2i\eta_{+\uparrow}\eta_{-\uparrow}\frac{2}{2\pi v_F}\eta_\sigma
J_{\perp F} J_{\perp B}\frac{d\lambda}{\lambda}
\sin(\sqrt{2}\phi_\rho(\tau))\cos(\sqrt{2}\phi_\sigma(\tau)).
\end{eqnarray}

   The last term to be considered in the second order cumulant expansion is the $S_z^2J_{zF}^2$ term.
 Scaling of this term,
however, does not contribute to the magnetic interaction and, also, does not produce any relevant
term for non-magnetic interaction. Collecting the results together, we now have the scaling
equations to the second order;

\begin{eqnarray}
\frac{dJ_{zF}}{dl}=\hspace{33mm}&&\frac{1}{2\pi v_F}
\left(\frac{1}{2\eta_\sigma}\left(\frac{1}{\eta_\sigma}+\eta_\sigma\right)
J_{\perp F}^2
+\frac{1}{2\eta_\sigma}\left(\frac{1}{\eta_\sigma}+\eta_\rho\right)
J_{\perp B}^2\right), \nonumber \\
\frac{dJ_{zB}}{dl}=\left(1-\frac{\eta_\sigma-\eta_\rho}{2}\right)J_{zB}
&+&\frac{1}{2\pi v_F}2\eta_\sigma J_{\perp F}J_{\perp B},\\
\frac{dJ_{\perp F}}{dl}
=\left(1-\frac{\eta_\sigma+\frac{1}{\eta_\sigma}}{2}\right)J_{\perp F}
&+&\frac{1}{2\pi v_F}(J_{zF} J_{\perp F}+\eta_\rho J_{zB}J_{\perp B}),\nonumber \\
\frac{dJ_{\perp B}}{dl}
=\left(1-\frac{\eta_\sigma-\eta_\rho}{2}\right)J_{\perp B}
&+&\frac{1}{2\pi v_F}(J_{zF}J_{\perp B}+\eta_\sigma J_{zB}J_{\perp F}).\nonumber
\end{eqnarray}
These equations can be simplified assuming the SU(2) symmetry for
the conduction electrons ($\eta_\sigma=1$);

\begin{eqnarray}
\frac{d(\rho_0 J_{zF})}{dl}=\hspace{37mm}& &(\rho_0 J_{\perp F})^2
+\left(\frac{1+\eta_\rho}{2}\right)(\rho_0 J_{\perp B})^2 \nonumber \\
\frac{d(\rho_0 J_{zB})}{dl}=\left(\frac{1-\eta_\rho}{2}\right)(\rho_0 J_{zB})
+& &2(\rho_0 J_{\perp F})(\rho_0 J_{\perp B})\\
\frac{d(\rho_0 J_{\perp F})}{dl}=\hspace{37mm}& &(\rho_0 J_{zF})(\rho_0 J_{\perp F})
+\eta_\rho(\rho_0 J_{zB})(\rho_0 J_{\perp B}) \nonumber \\
\frac{d(\rho_0 J_{\perp B})}{dl}=\left(\frac{1-\eta_\rho}{2}\right)(\rho_0 J_{\perp B})
+& &(\rho_0 J _{zF})(\rho_0 J_{\perp B})+(\rho_0 J_{zB})(\rho_0 J_{\perp F}), \nonumber
\end{eqnarray}
where $\rho_0(=1/{2\pi v_F})$ is proportional to the density of
state. As we see later, for ferromagnetic J, the triplet ground state can be formed due to
 the interacton parameter $\eta_\rho$ in the second order term. In 1D correlated
electron system, the electron-electron interaction induces a short range magnetic
ordering which, in turn, introduces a molecular field on the magnetic impurity\cite{Fulde}.
Therefore,  the impurity and a electron form a triplet aligned to the molecular field, thus,
 resulting  the broken local SU(2) symmetry.  These equations are naturally reduced to the
scaling equations of the conventional Kondo model when there is no
Luttinger interaction i.e. $\eta_\sigma=1$ and $\eta_\rho=1.$

So far, no assumption has been made on the strength of
$\eta_\rho(=[(1-g/\pi v_F)/(1+g/\pi v_F)]^{\frac12})$. Here, g is
the strength of scattering between the right-going and the
left-going mode and the scattering within the same mode is
neglected. In the small g regime, the above scaling equations are
reduced as follows,
\begin{eqnarray}
\frac{d(\rho_0 J_{zF})}{dl}=\hspace{29mm}
   & &(\rho_0 J_{\perp F})^2+(\rho_0 J_{\perp B})^2
   -\frac{g}{2\pi v_F}(\rho_0 J_{\perp B})^2 \nonumber \\
\frac{d(\rho_0 J_{zB})}{dl}=\frac{g}{2\pi v_F}(\rho_0 J_{zB})
   +& &2(\rho_0 J_{\perp F})(\rho_0 J_{\perp B})\\
\frac{d(\rho_0 J_{\perp F})}{dl}=\hspace{29mm}
   & &(\rho_0 J_{zF})(\rho_0 J_{\perp F})+(\rho_0 J_{zB})(\rho_0 J_{\perp B})
   -\frac{2g}{2\pi v_F}(\rho_0 J_{zB})(\rho_0 J_{\perp B}) \nonumber \\
\frac{d(\rho_0 J_{\perp B})}{dl}=\frac{g}{2\pi v_F}(\rho_0 J_{\perp B})
   +& &(\rho_0 J_{zF})(\rho_0 J_{\perp B})
   +(\rho_0 J_{zB})(\rho_0 J_{\perp F}). \nonumber
\end{eqnarray}
We note that the above result is in agreement with the FN's poor
man's scaling result except for the symmetry breaking terms involving
the g parameter. In fact, the extra terms in our RG equations
correspond to the next higher order terms which were neglected in
the FN approach.

 \section{The Ground State}

 The scaling equations yield two strong coupling fixed points;
   $(J_F,J_B)= (\infty,\infty)$ and $(\infty,-\infty)$. The
first fixed point governs the antiferromagnetic regime, which gives the singlet
 as the ground state. The second one corresponds to the ferromagnetic coupling.
 However it is not clear whether this fixed point corresponds to
  a singlet or to a triplet state.
We have calculated  flows of the coupling constant for several values of the Luttinger
 interaction parameter, $\eta$, for the ferromagnetic fixed point.
  For finite $\eta$ other than unity, $J_B/J_F < -1$ and
 $J_B+J_F$ flows to $-\infty$, whereas  $J_B+J_F$ is equal to zero
 when $\eta=1$(Fig. 1). When the coupling constant $J$ grows and becomes
 large, the Kondo coupling term becomes dominant in Eq.(35). Therefore,
 the Luttinger
 interaction becomes irrelevant\cite{FN} and we can treat
 the impurity spin as a classical spin with the magnitude $S=1/2$.

The asymtotic Hamiltonian which represents this situation can be written as
\begin{eqnarray}
H=\sum_{rk\sigma}\epsilon_k c^{\dagger}_{rk\sigma}c_{rk\sigma} +\frac14\sum_{rk,r'k',\sigma}
J_{rr'\sigma}c^{\dagger}_{rk\sigma}c_{r'k'\sigma'},
\end{eqnarray}
where $J_{11 \sigma}=J_{22 \sigma}=\sigma J_F$ and $J_{12 \sigma}=J_{21 \sigma}=\sigma J_B$. The
Luttinger interaction effect is included in the coupling constant
 through the renormalization process. The corresponding Green's functions
are given  as follows ;

\begin{eqnarray}
G_{11}(\epsilon)&=&G_{22}(\epsilon)
 =G_0(\epsilon)+G_0(\epsilon) \frac{\frac14 \sigma J_F}
 {1-\frac14 \sigma(J_F+J_B)G_0(\epsilon)}G_0(\epsilon), \nonumber\\
G_{12}(\epsilon)&=&G_{21}(\epsilon)
 =\hspace{15mm}G_0(\epsilon)\frac{\frac14 \sigma J_B}
 {1-\frac14 \sigma(J_F+J_B)G_0(\epsilon)}G_0(\epsilon).
\end{eqnarray}

For $\eta=1(J_F+J_B=0)$, there exist no bound state since the Green's functions do not have poles.
This case has the same properties as the three-dimensional ferromagnetic Kondo coupling case.
However, if the Luttinger interaction is turned on$(\eta<1)$, $J_F+J_B$ flows to $-\infty$. The
Green's functions have a pole for $\sigma=+1$ suggesting that a conduction electron becomes
bounded \cite{Doniach} to form a triplet state in the ground state for the ferromagnetic Kondo
exchange.

\section{The Scalar Potential Scattering Effect}

In a more realistic model, a magnetic impurity generates
scattering due to an elastic potential, $\omega$\cite{EK}.
Recently, Wang and his coworkers have carried out  Bethe ansatz
calculations on the one-dimensional Kondo problem\cite{WV,WE}.
They showed that for an attractive potential scattering, the bound
state always appears, whereas for a repulsive potential
scattering, there exists a bound state for $|J|>4\omega$, but no
bound state for $|J|<4\omega$.

  The scaling theory for the scalar potential in the first order is
similar to the magnetic interaction. The bosonized term is

\begin{eqnarray}
\frac{1}{2\pi\alpha}(2\omega_F\alpha\partial_x\sqrt{2}\phi_\rho(0)
+4i\omega_B\eta_{+\uparrow}\eta_{-\uparrow}
\cos(\sqrt{2}\phi_\rho(0))\sin(\sqrt{2}\phi_\sigma(0))),
\end{eqnarray}
 and we have the scaling equations

\begin{eqnarray}
\frac{d\omega_F}{dl}=0,\hspace{10mm}
\frac{d\omega_B}{dl}=\frac{1-\eta_\rho}{2} \omega_B,
\end{eqnarray}
showing that the backscattering contribution is relevant. In the
second order, the candidates for renormalizing the forward
potential scattering are contributions from the terms like
$\omega_F^2$ or $\omega_B^2$. But, we find that $\omega_F^2$
yields only a constant by a simple calculation,
\begin{eqnarray}
\frac{1}{(2\pi\alpha)^2}4\omega_F^2\alpha^2\int d\tau\int d\tau'
<\partial_x\sqrt{2}\phi_{\rho_f}(\tau)\partial_x\sqrt{2}\phi_{\rho_f}(\tau')>,
\end{eqnarray}
which clearly has no slow mode. Similarly, the $\omega_B^2$ term generates terms such as
$\cos(2\sqrt{2}\phi_\sigma(\tau))$, and $\cos(2\sqrt{2}\phi_\rho(\tau))$, which are irrelevant as
discussed above. Also, we can show that $J_{zF}^2$, $J_{zB}^2$, $J_{zF}J_{zB}$, $w_Fw_B$ and
$J_{z\perp,FB}\omega_{FB}$ are also irrelevant through the same procedure as in $\omega_F^2$ and
$\omega_B^2$. Thus, the potential scattering is not scaled at the second order cumulant and there
is no interplay between the magnetic and the non-magnetic interaction. Actually, this situation
is not unexpected, since we know that the nearest neighbor coulomb interaction in the Hubbard
model does not affect the magnetic properties\cite{Auerbach}.

In the case of an impurity spin, S=1/2, the energy of a single electron coupled with the impurity
is such that the ferromagnetic coupling energy is $\frac{1}{4}J$ and the scalar potential is
$\omega$. It means that if $4\omega>|J|$ initially, the potential energy grows infinitely along
the scaling process so that no bound state can be formed. If $4\omega<|J|$ initially, a spin
triplet state is formed. Under the open boundary condition, this corresponds to a  bound state of
a spin $\frac{3}{2}$ complex in agreement with the Bethe ansatz results\cite{WV}. Here, we note
that, in the present treatment, it is not necessary to invoke the nearest neighbor spin
interaction as done by FN. We believe that the short-ranged nature of the spin interaction makes
the FN scenario unlikely, although it cannot be ruled out completely.

\section{The Kondo Temperature}

In the conventional three-dimensional Kondo model, the Kondo temperature is given  by $T_K=D
e^{-\frac1{2J}}$, which originates from the scaling equation $dJ/dlnD=-2J^2$, where D is the
bandwidth. In their previous study, Lee and Toner showed that the scaling relation,
$dJ_B/dlnD=-\frac{1-\eta}2 J_B$, gives a power law Kondo temperature, $T_K=D
J_B^{\frac2{1-\eta}}$\cite{LT}. However, the crossover behavior from the Fermi liquid to the
Luttinger one has not been studied. In order to address this question, we consider the scaling to
second order for the forward and the backward scattering simultaneously.

From the RG flow, we get the coupling constants as  functions of
$D$, i.e. $\tilde{J_F}(\tilde{D})$ and $\tilde{J_B}(\tilde{D})$,
where $\tilde{J_F}$ and $\tilde{J_B}$ are the scaled couplings and
$\tilde{D}$ is the scaled bandwidth cutoff. The Kondo temperature
is an invariant energy scale in the scaling procedure and, thus,
can be expressed as $T_K=\tilde{D}f(\tilde{J})$, where
$f(\tilde{J})$ becomes exponential  or power law type depending on
the limiting case. For some initial value of $J$, $T_K$ remains
constant through the scaling procedure. In such a case,
$f(\tilde{J})$ becomes proportional to
$\frac1{\tilde{D}(\tilde{J})}$, thus yielding the Kondo
temperature.

$T_K$ for the backward scattering part is given in Fig. 2. It can be clearly seen that $T_K$ is
exponential for a weak Luttinger interaction and
 becomes  power law type as the Luttinger interaction
increases. From inset, we observe that the linear  slope of linear plot$(\eta=1)$ is -1/2 in
agreement with $T_K=De^{-\frac{1}{2J}}$.

\section{Summary}

  In summary,  we have studied the Kondo effect in a Luttinger
Liquid in presence of a scalar potential. We have obtained the scaling equations up to the second
order for an arbitrary Luttinger interaction strength by renormalizing the boson fields.

The ferromagnetic fixed point is studied using an asymptotic Hamiltonian. It is shown that a
triplet bound state can be formed in agreement with the recent Bethe ansatz calculation without
invoking the nearest neighbor spin interaction. The Luttinger  interaction induce the triplet state to break
 the local SU(2) symmetry  about the impurity spin contrary to the result obtained by Furusaki and Nagaosa.
 The magnetic interaction and the potential
scattering do not interplay and the triplet state is sustained for a weak scalar potential,
$|J|>4\omega$. The Kondo temperature for arbitrary strength of the Luttinger interaction is
calculated. The result shows a clear crossover behavior from an exponential to a power law type.

$\bf{Acknowledgement}$

This work is partially supported by the Korea Research Foundation(99-005-D00011).

$\\$

$\bf{Figures}$

Fig. 1 The coupling constant $J_{F,B}$ for several Luttinger interaction strength $\eta$. The
magnitude of slope and the value of $|J_F+J_B|$ become larger as the interaction strength grows
larger. The slope of the dotted line is -1.

Fig. 2 The Kondo temperature as  a function of the backward scattering coupling constant, $J_B$,
for several Luttinger interaction strength $\eta$, where $T_K^o$ is the Kondo temperature for
each $\eta$ at $J_B/D=0.1$. The linear to nonlinear crossover is clearly shown. Note the log-log
scale of the graph. The inset shows the Kondo temperature as a function of $J_B^{-1}$. The linear
curve for $\eta=1$ clearly shows the $T_K=De^{-\frac1{2J}}$ behavior.
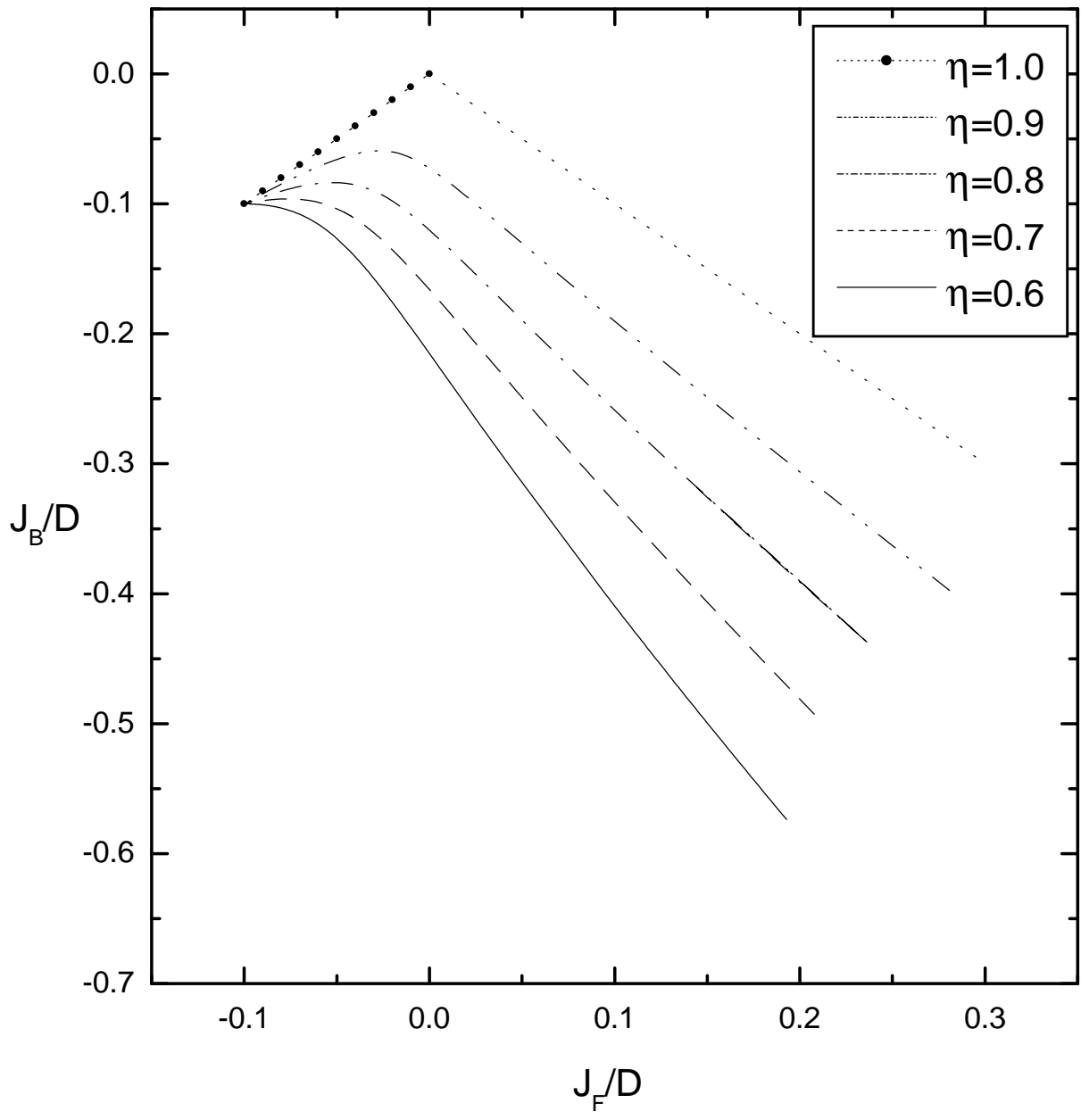
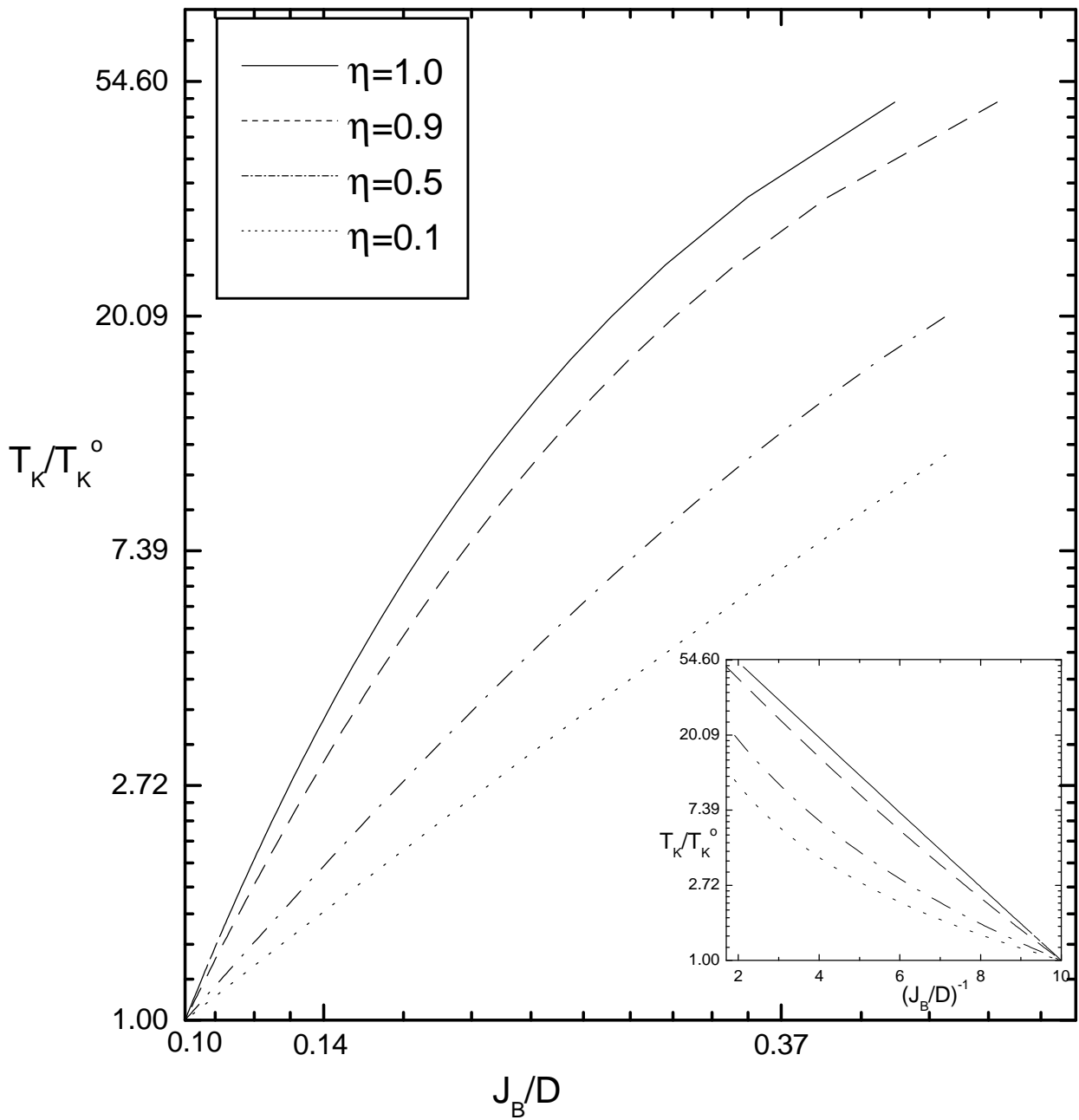
\end{document}